\documentstyle[prl,aps]{revtex}
\begin{document}
\twocolumn[
\hsize\textwidth\columnwidth\hsize\csname@twocolumnfalse\endcsname
\draft
\title{Spin-Charge Separation in Angle-Resolved Photoemission Spectra}
\author{Hidekatsu Suzuura and Naoto Nagaosa}
\address{
Department of Applied Physics, University of Tokyo,
Hongo, Bunkyo-ku, Tokyo 113, Japan}

\date{\today}
\maketitle
\begin{abstract}
The implications of the angle-resolved photoemission spectroscopy
(ARPES) on the issue of  spin-charge separation are discussed from the 
viewpoint of slave-particle RVB theories. It is pointed out that the 
non-local phase string is essential to 
reproduce the two-peak structure experimentally observed in 
SrCuO$_2$.
The comparison between 1D and 2D cases is also made.
\end{abstract}
\pacs{71.27.+a, 78.20.Bh, 79.60Bm.}
]
The spin-charge separation is one of the central issues in the physics 
of strongly correlated electronic systems. 
Experimentally the angle-resolved photoemission spectroscopy
(ARPES) became a powerful tools to investigate this issue,
because it gives the information on the 
electron Green's function directly as a function of the 
momentum and energy.
Recently Kim et al. reported ARPES in one-dimensional SrCuO$_2$
\cite{shen},  which is a typical example of a half-filled Mott insulator.
In the spectra the two peak-structures with different dispersions are
observed, and they interpreted the results as the dispersions of 
spinon and holon
and hence the first experimental evidence for the spin-charge separation in 1D.
However the word ``spin-charge 
separation'' is not yet uniquely defined, and especially the 
relation between spin-charge separation in 1D and 2D remains 
unclear. The definition of the spin-charge separation we employ in this paper
is that the
electron operator $(c_{i \sigma})$ is given
by the product of the spinon $(s_{i \sigma})$ and holon
$(h^{\dagger}_{i \sigma})$ operators as \cite{anderson}
\begin{equation}
c_{i \sigma} = s_{i \sigma} h^\dagger_i,
\end{equation}
where both spinon and holon can be basically
regarded as free bosons or fermions
with definite momentum-energy relation.
In this case the problem is reduced to that of one-particle
properties. 
A RVB mean-field Hamiltonian of the $t$-$J$ model becomes
\begin{equation}
H=-\frac{t_h}2 \sum_{<i,j>} h_i^{\dagger}h_{j} + 
\frac{J_s}2 \sum_{<i,j>,\sigma} s_{i,\sigma}^{\dagger}s_{j,\sigma} 
\end{equation} 
In this paper we argue that the spin and charge are {\it not}
separated in their experiments by the definition given above.
In contrast it will be shown below 
that the peak which they attributed to the holon
is the evidence for the non-local phase string connecting them.

The physical picture for eq.(1) is that, when one photo-hole is created by the
incident light, one vacant site (holon) and one (anti-)spinon
are created. In 1D the spinon can be regarded as the kink of the
antiferromagnetic configuration and can move freely without any
energy cost proportional to the distance $L$ from the holon.
Hence the holon has the dispersion
$E^{\rm holon}_k = -t_h \cos k$ while the spinon
$E^{\rm spinon}_k = -J_s \cos k$.
If this naive picture is correct, 
there are energy-momentum relations
given by
\begin{eqnarray}
k  &=& k_s - k_h
\nonumber \\
\omega  &=& 
E^{\rm spinon}_{k_s} -
E^{\rm holon}_{k_h}.  
\end{eqnarray}
where $k$ and $\omega$ is the momentum and energy of the electron.
Here, we adopt the slave-boson formalism.
In this representation the holon is a boson and the spinon is a
fermion. At half-filling, there is no hole and the statistics of holon 
has no effect on the spectral function. On the other hand, the spinons form the
half-filled Fermi sea with the Fermi level $E=0$ and this is essential 
to the following results.

The electron Green's function,
which is observed in ARPES, 
is given in real space-time by the product
of those of spinon and holon as
$G(r,t) = G_{\rm spinon}(r,t) G_{\rm holon}(-r,-t)$\cite{lee}.
The Fourier transform  $G(k,\omega)$ is given by the convolution,
and the spectral function 
$A(k,\omega) = - \pi^{-1} {\rm Im} G(k,\omega)$ is
\begin{eqnarray}
& &A(k,\omega) \nonumber \\
&=&\frac1{2\pi}\int_{-\pi}^{\pi}d{k_h} 
\int_{-k_s^F}^{k_s^F} dk_s \delta(k+k_h-k_s) \delta(\omega+ E^{\rm
holon}_{k_h}  
-E^{\rm spinon}_{k_s}) \nonumber \\
&=& \frac1{2\pi}\frac1{\sqrt{F_k^2-\omega^2}} f(k,\omega).
\end{eqnarray}
The momentum cut-off $k_s^F$ in the $k_s$-integration determines the
support $f(k,\omega)$ of the spectral function, and other symbols are
defined by
\begin{eqnarray}
f(k,\omega) &=&
\left\{ \matrix{\theta(F_k-\omega)\theta(\omega+F_k\sin \phi_k) &
\left(0\le k \le k_0 \right) \cr
\theta(F_k\sin \phi_k-\omega)\theta(\omega+F_k) &
\left(k_0 \le k \le \pi \right)}\right., \nonumber \\
F_k&=& \sqrt{t_h^2+J_s^2-2t_h J_s \cos k}, \nonumber \\
\tan \phi_k &=&\frac{t_h \sin k}{t_h \cos k  -J_s}, \ \ 
\phi_{k_0}=\frac{\pi}{2},
\end{eqnarray}
where $\theta(x)$ is the Heaviside step function. 
We assume that $\phi_k$ becomes $k$ in the limit $J \to 0$.
The fully shaded line shape in Fig.1(a) shows the $A(k, \omega)$ for
$k=\frac{\pi}6$. 
The line shape is almost determined by the DOS of holon and the
momentum-dependent cut-off energy due to the filled Fermi sea of
spinon.
The square-root singularity appears at the band edge of holon and 
the energy dispersion of spinon gives the $F_k$ with the
band width $2 J_s$. The cut-off energy has the dispersion $t_h$,
but the peak structure with such a dispersion is missing here.

What is needed additionally to reproduce the correct behavior
is the {\it phase string}  effect which has been discussed by Weng et al.
\cite{weng}. After the unitary transformation to take care of the
sign change of the wavefunction, the
electron operator is given by
\begin{equation}
 {\tilde c}_{i \sigma} =  {s}_{i \sigma} {h}^{\dagger}_i
e^{i (\theta_i^h+\theta_i^s)} \label{semi} 
\end{equation}
with
\begin{eqnarray}
\theta_i^h &=& \mp { \pi\over 2}  \sum_{l>i} {h}^{\dagger}_{l}{ h}_{l} 
\nonumber \\
\theta_i^s &=& \pm \frac{\pi}2   \sum_{l>i} ({s}^{\dagger}_{l
\uparrow} {s}_{l \uparrow}+{s}^{\dagger}_{l\downarrow} {s}_{l
\downarrow} -1) \nonumber 
\end{eqnarray}
Here, the spinon and the holon are described by the {\it free fermions}
${s}_{i \sigma}$ and ${h}_i$, respectively, but 
the original spinon and holon
strongly interact with each other via nonlocal phase-string.
Now that the system is at half filling, the phase-string term of holon 
has no contribution, but that of the spinon makes the long-distance behavior
of the spinon Green's function change 
from $G_{\rm spinon}(x,t)\sim e^{ \pm i k_s^F x}/(x\pm v_s t)$ to 
\begin{eqnarray}
G_{\rm spinon}(x,t)&=&\langle s_{x,\sigma}^{\dagger} e^{-i\theta_x^s}
e^{i\theta_0^s} s_{0,\sigma} \rangle \nonumber \\
&\sim& \frac{e^{ \pm i k_s^F x}}{\sqrt{x\pm v_s t}},
\end{eqnarray} 
where the Fermi velocity of spinon $v_s=dE^{\rm spinon}_{k=k_s^F}/dk=J_s$.
The bosonization technique gives this asymptotic behavior under the
constraint $\langle {s}^{\dagger}_{l\uparrow}{s}_{l
\uparrow}+{s}^{\dagger}_{l\downarrow} {s}_{l\downarrow} \rangle=1$
\cite{weng}. 

Taking this asymptotics into account, the spectral function $A(k,
\omega)$ is given by 
\begin{eqnarray}
& &A(k, \omega) \nonumber \\ 
&\sim& \int dx dt dk_h e^{i(\omega+E^{\rm holon}_{k_h})t
-i(k+k_h\mp k_s^F)x} \frac1{\sqrt{x\pm v_s t}} \nonumber \\
&\sim& \int dX dk_h {e^{-i(k+k_h\mp k_s^F)X}}\frac1{\sqrt{X}}
\nonumber \\  \ \ \ & & \times
\delta\left(\omega+E_{k_h}^{\rm holon}\mp v_s(k+k_h\mp k_F^s)\right).  
\end{eqnarray}
This result is valid when low-energy anti-spinons are
created below the Fermi points, that is, $\Delta k=k_h+k\mp k_F^s$ and
$\Delta\omega=\omega+E_{-k\pm k_s^F}^{\rm holon}(>0)$ are close to zero.
The constraint originated from the $\delta$-function is expanded in
terms of $\Delta k$,
\begin{eqnarray}
& &\omega+E_{k_h}^{\rm holon}\mp v_s(k+k_h\mp k_F^s) \nonumber \\
&=& \Delta\omega\pm(t_h \cos k-J_s)\Delta k 
\mp \frac12 t_h \sin k (\Delta k)^2+O\left( (\Delta k)^3 \right)  \nonumber \\
&=& 0. \nonumber
\end{eqnarray}
For $k \ne k_0$, $\Delta \omega$ is in proportion to $\Delta k$.
So, $\delta$-function gives no singularity and we obtain  
\begin{equation}
A(k, \omega)\sim \int dX \frac{e^{-i\Delta k
X}}{\sqrt{X}}\sim (\Delta k)^{-\frac{1}{2}} \propto
(\Delta\omega)^{-\frac12}.  
\end{equation}
Then the square root singularities are restored.
For $k=k_0$, $\Delta\omega$ scales as $(\Delta{k})^{2}$
and $\delta$-function gives a singular contribution 
$\sim 1/\Delta k$, which is identical to the band edge singularity of
holon. So, the critical exponent changes as
\begin{equation}
A(k, \omega)\sim (\Delta k)^{-\frac12-1} \propto
(\Delta{\omega})^{-\frac34}.  
\end{equation}
The same exponents in 
the limit $U \to \infty$ are 
obtained by Sorella and Parola \cite{sor}.
That is the case also for the finite $J$ in this approximation, which
disagrees with the conclusion of them who claimed that the singularity 
becomes $(\Delta \omega)^{-\frac12}$ everywhere for finite $J$.
The partially-shaded portion in Fig.1 (a) schematically shows the 
singular structures due to the phase string interactions. 
As mentioned before, one anti-spinon is
created, in other words, one spinon is annihilated 
under the photoemission process. Thus the energy
tail goes at the positive-energy side.

In Fig.1 (b) the peak positions of the spectral
function are plotted in $\omega$-$k$ plane. 
We interpret these structures as follows for small
$J_s$ case. When one photoelectron emits, one holon is created.
Without spinon excitations, a single quasi-particle peak with the
dispersion $t_h$ is observed. One might think that the experimental spectra
can be interpreted by adding the peak structure with the dispersion
$J_s$ due to the spinon excitations. However, the situation is not so
simple.    
If spinon excitations are incoherent, the momentum conservation breaks
and the whole DOS of holon is observed in the spectra. 
The square root singularity due to the DOS of holon appears at the
band edge and has the dispersion $J_s$ due to the spinon.
Actually the spinon is an excitation of the coherent spin liquid
forming the half-filled Fermi sea. So the
energy and momentum conservation make spinon
annihilation unable over the Fermi sea and this
gives $k$-dependent energy cut off in the spectral function. 
What is more, the phase-string interactions causes 
the divergent structures like Fermi edge singularities 
concerned with low-energy excitations near the Fermi level. 
Energy-momentum conservation makes 
these singularities located at the holon energies with the momentum
shift $-k$ from the two Fermi points of spinon and 
$\omega_s=t_h \cos(\pm k_F^s-k)=\pm t_h \sin k$. 
Thus they have the dispersion $t_h$. 
The right side of this Fig.1 (b) agrees well with the
results of the experiment and the exact diagonalization
calculation\cite{shen}.   
Therefore the two peak structure with the dispersion $t_h$ can
be interpreted as the evidence not for the noninteracting spinpon and holon
but for the nonlocal phase string connecting them.

The above spectral function has been already obtained by the more elaborated
treatment by Sorella and Parola \cite{sor} and Penc et al. \cite{penc}
based on the Bethe-Ansatz wavefunction
given by Ogata and Shiba \cite{ogata} in the $U \to \infty$ limit, which
is the product of spin and charge parts. 
The phase string effect is already built-in in this wavefunction.
However it was not clear whether the slave-particle formalism can 
reach the correct answer,
and the calculation above gives a simple approximate means to reproduce
it starting from eq.(1).

The issue of spin-charge separation is related to the {\it phase string},
i.e., the quantal phase, described above.
In the gauge theoretical formulation of the $t$-$J$ model
we identify the problem of spin-charge separation as the 
confinement-deconfinement problem \cite{anderson,lee,nagaosa}.
In the lattice gauge theory it is known that in 
1D the deconfining phase, i.e., spin-charge separated phase, is
impossible \cite{kogut}.
Then the gauge theory predicts that the naive spin-charge separation 
does not occur in 1D which is consistent with the 
correct answer. We believe that this gives some credit to the 
slave-particle + gauge theoretical formalism.
Therefore the experiment of Kim et al. \cite{shen} is the evidence for the
spin-charge separation in the sense of Bethe-Ansatz
wavefunction, but not in the sense of eq.(1).

It is noted that the slave-fermion formalism can give the correct 
behavior. In that representation, the holon is a fermion and the
spinon is a boson. This is closely related to the fact that, at
half-filling, the Bethe-Anzats wavefunction is a product of a
Slater-determinant of spinless fermions and the spin wavefunction of
the Heisenberg model. The holon is certainly a spinless fermion
in the slave-fermion formalism and the Heisenberg spin can be
described by the Schwinger-boson representation.  
In order to reproduce the correct behavior, however, the free-boson
approximation is not enough and the hard-core condition must be taken
into account. With this interaction between spinons, the correlation
function $\langle b_{x\sigma}^{\dagger} 
b_{0\sigma} \rangle$ scales as $x^{-\frac12}$ in the long-distance limit, 
and this behavior corresponds to that of the Bethe-Anzats solution in
the limit $U \to \infty$\cite{ogata}.
The hard-core interaction between the bosons is automatically taken
into account in terms of free fermions with the $\pi-$flux
phase-string, {\it i.e.}, Jordan-Wigner transformation
\begin{equation}
b_{i}=f_{i} e^{i\pi\sum_{l>i}f_l^{\dagger}f_l}. \nonumber
\end{equation}  
This again strongly suggests the importance of the phase string effects.
Recently, it is argued that a semionic-decomposition is an appropriate
representation to grasp the key physical properties of the $t$-$J$
model \cite{mar}. 
 It is certain that eq.(\ref{semi}) can be interpreted as a semionic
decomposition and the free semion theory gives correct behavior.
However, there is no one-particle basis for free semions at the
moment, and hence the free-semionic decomposition does not satisfy our 
criterion of the spin-charge separation.
Free semions strongly feels infinite-range
phase-string interactions although the Hamiltonian has no interaction
terms. 

The real purpose of the slave particle formalism is of course to 
study the higher dimensional systems.
In higher dimensions, there occurs the  energy increase 
proportional to the separation 
$L$ between the spinon and holon if the spin configuration is static.
This {\it energy string}, however, disappears when 
the holon waits for the spins to flip to
relax the energy cost, and the holon dispersion becomes of the order of
$J$\cite{kane}. This explains the ARPES  in
Sr$_2$CuO$_2$Cl$_2$ (2D) where only one peak is observed
with the dispersion width $\sim J$.
However this does not necessarily mean that the spin and charge are confined,
because the confinement is the issue of the phase string.
In (2+1)D it has been discussed that the 
deconfining phase is possible 
with the dissipative dynamics of the gauge field \cite{nagaosa},
which is in contrast to the energy string consideration given above.
In the Neel state with long range magnetic ordering, the 
gauge field is confining with the chiral symmetry breaking.
This corresponds to the undoped
Sr$_2$CuO$_2$Cl$_2$ (2D), and the ARPES can be basically 
understood  in terms of the spin density wave picture
with $U\gg t$.
When the holes are doped,  the Neel order disappears and the
system becomes conductive. It has been pointed out that the 
gauge field may be deconfining in this case \cite{nagaosa}.
If the gauge field is deconfining,
the mean field theory in the slave boson formalism \cite{fukuyama}
is basically justified with the perturbative corrections
due to the gauge field fluctuations.
The spectral function has only one peak in this case \cite{lee}.
Experimentally $A(k,\omega)$ at finite doping
has only one peak but seems to consist of two components,
i.e., the broad background and the quasi-particle peak \cite{shen2}.
The latter is not resolved from the former in the normal state,
but it becomes sharp below the superconducting transition temperature $T_c$.
Additionally the weight of the quasi-particle peak
grows as the doping $x$ increases \cite{shen2}.
These features are consistent with the
prediction assuming the naive spin-charge separation in eq.(1) \cite{lee}.

In conclusion we interpret the two-peak structure in ARPES of 
SrCuO$_2$  as the evidence for the phase string effect. This is 
in contrast to the two-dimensional high-Tc cuprates
where the spectral function is composed of two components but has only one 
peak, which is consistent with the spin-charge separation in the slave boson 
theory. The spin-charge separation is related not to
the {\it energy string}  but to the more subtle {\it quantal phase string}.

The authors acknowledge S.Maekawa, K. Pence, and H. Shiba for useful
discussions. 
This work is supported by Priority Areas Grants 
and Grant-in-Aid for COE research
from the Ministry of Eduction, Science and Culture of Japan,

\begin{figure}
(a) The spectral function $A(k,\omega)$  with $k=\frac{\pi}6$. 
The simple slave-boson decomposition gives the fully-shaded line
shape. The partially-shaded singularities are due to the nonlocal
phase-string interactions.
(b) The locations of the singularities are ploted in the $\omega$-$k$
plane. 

\end{figure}
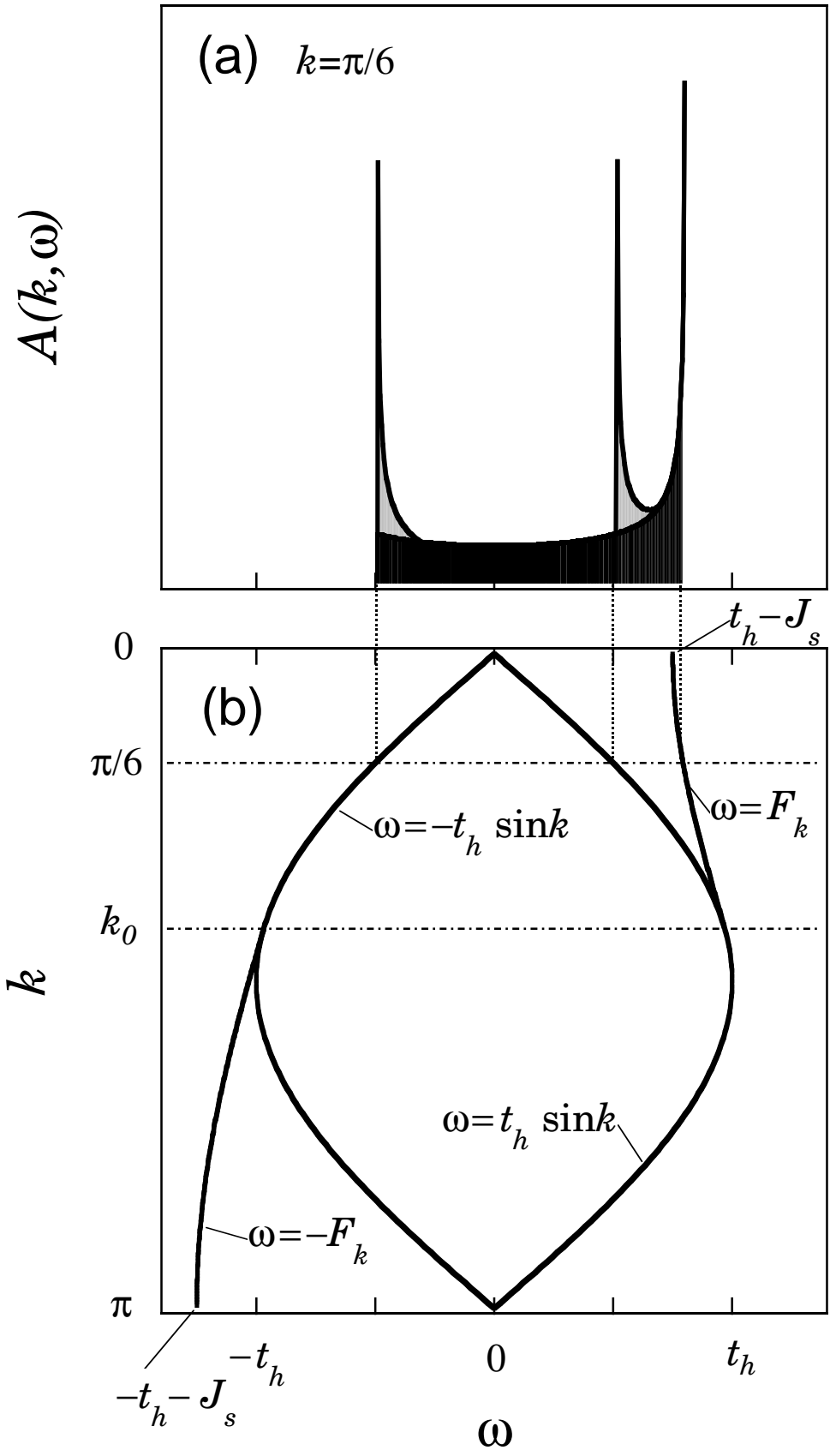
\end{document}